# A Basic Model of KBS Software

A. N. Clark, Computer Systems Division, GEC-Marconi Research,
Great Baddow, Chelmsford, Essex, CM2 8HN, Great Britain
e-mail: anc@gec-mrc.co.uk, tel.: (0245) 473331

This document will describe a simple model of conventional software development, show how KBS software development differs from the conventional and then define a collection of terms which are important with respect to the quality of the software under development. We will be deliberately vague about the exact scope of the term *software development*; it will certainly cover the activity of implementation, but it is also intended to cover some aspects of software specification and design.

The term KBS software is very difficult to pin down, as certain features which are argued to be characterising, such as *knowledge bases*, *rules* and *symbolic representation*, turn out to be difficult to define in any meaningful sense or are not characterising at all. The definition which is used in this document views the characterising feature of KBS software as being defined by a *non-deterministic* operator and its affect on the behaviour of the program which is being developed.

In order to define a model of software development, it will be necessary to keep the notion of a program simple and precise. The $\lambda$-calculus has been chosen as the class of programs which will motivate the discussion. This has the advantage of being small, can be given a simple semantics and captures many classes of programming language.

The rest of this document is divided into the following sections:

- A simple model of program development is described in terms of the $\lambda$-calculus and the SECD machine.

- The characteristic feature of KBS software is described and its impact on software development is discussed.

- A collection of terms which affect software quality are defined with respect to the model of software development.

The presentation will be kept informal but precise, we have in mind a formal underlying presentation of these ideas which can be developed if the basic notions turn out to be useful.

## 1 Software development

A *program* will be defined to be an expression in the $\lambda$-calculus whose concrete syntax is defined as follows:

$$E ::= I \mid \lambda I.E \mid E_1 E_2$$

where $E$ is to be thought of as the set of all programs. Such a syntactic definition states nothing about the meaning of each program. One way of giving the programs a meaning is by defining *how* they will be performed using a transition *machine*. Different machines will attribute different meanings to the programs by encoding specific builtin evaluation mechanisms and types of value. This will be the way in which the programs will be given a meaning in software development.

Here is the basis for all machines which treat the λ-calculus as being *call-by-value*:

$$(s, b, i :: c, d) \longmapsto ((b \bullet i) :: s, b, c, d)$$

$$(s, b, (\lambda i.e) :: c, d) \longmapsto (<i, b, e> :: s, b, c, d)$$

$$(s, b, (e_1 e_2) :: c, d) \longmapsto (s, b, e_2 :: e_1 :: @ :: c, d)$$

$$(<i, b_1, e> :: v :: s, b_2, @ :: c, d) \longmapsto ([], b_1 \oplus (i \mapsto v), [e], (s, b_2, c, d))$$

$$(v :: \_, \_, [], (s, b, c, d)) \longmapsto (v :: s, b, c, d)$$

The SECD machine is described in detail elsewhere [9], the following is a brief description. $s$ is a list of program outcomes called the stack, $b$ is a collection of bindings between program identifiers and outcomes called the environment, $c$ is a list of programs, $e \in E$, and machine instructions @ called the control and $d$ is a machine state called the dump. Program identifiers, $i \in I$, are looked up in the environment $b$ using $b \bullet i$, when a λ-expression, $\lambda i.e$, is performed the result is a *closure* $<i, b, e>$ which may be applied to a value $v$ in which case the body of the closure is evaluated with respect to an environment which is $b$ extended with a binding between $i$ and $v$. During an application the current state is saved on the dump as a *resumption point* and will be resumed when the application is complete.

All manner of different machines can be defined by

- Changing some of the transition rules. For example we could produce a radically different language by changing how function calls are performed.

- Introducing some new data types. An *outcome* of the machine is a value which can occur at the head of the stack when the machine reaches a terminal state (one in which there is no control left). As it stands, the machine given above defines *closures* (written $<\_, \_, \_>$) as the only outcomes. New types of outcome may be added and some builtin operators which create and manipulate them. For example the machine can be enriched with integers and some builtin operators such as $+$ and $*$.

- Introducing some new transition rules. A new transition rule will perform some computation which is not possible with the current facilities. For example a builtin operator *halt* could be added which will cause the current computation to be aborted.

Some of these machines will correspond to concrete programming languages such as C or Ada but others will represent entirely new languages. The intention is that the machines are not exclusively restricted to describing a concrete implementation, but may be also used to describe aspects of a design and even aspects of a specification. This is possible because there are no restrictions on the type of computation which can occur due to the machine transitions. Such computations can be designed specifically for the job in hand and may be as abstract and high level or as concrete and low level as desired.

It is important to note that the machine computes by proceeding from state to state. Each state is a self contained unit and represents a snapshot of the computation at some time and each state proceeds from the previous state using the same type of indivisible transition. The entire evaluation which a machine will produce when a program is performed is represented as a complete sequence of the states which the machine will pass through from the initial state to the final state. Such a sequence will be termed a *calculation*.

The set $E$ contains all the possible programs which can be written. The development process can proceed in one of two ways

- Starting with an idealised representation of the desired system, changes are applied to the system in terms of modifications and extensions to the program or changes to the semantics of the program, until a concrete (*i.e.* implementable) program is produced. At each stage any change is carefully controlled so that it is possible to state *how* the result differs from the initial ideal.

- Starting with an approximation to the desired system and a collection of criteria for success, changes are applied until the system meets the criteria. It is understood exactly *why* the initial system does not meet the success criteria which will guide the changes which are necessary to meet them.

Both of the approaches rely on a clear understanding of what it means to *change* a program. We will define a change to a program to be one of the following:

- A *modification* which textually replaces part of a program whilst the language semantics remains the same.

- An *extension* which wraps some new program around an existing program whilst the language semantics remains the same.

- A *port* which involves translating from one program to another and changing the underlying semantics of the programming language.

Program modification will be referred to as *intra*-language changes whilst a port is an *inter*-language change.

The intra-language changes must respect the structure of the programs as defined by $E$. Each program is defined to be a term with 0 or more subterms. For example the following diagram represents a $\lambda$-function as a term where the triangle marked $e$ represents a complete subterm.

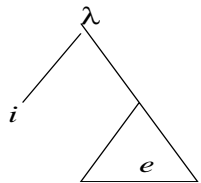

The following diagram represents an application with two subterms marked $e_1$ and $e_2$ respectively.

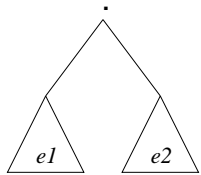

The other type of term defined by $E$ is the single identifier $i$ which contains no subterms. A modification involves taking a single program, such as the application term above, and replacing a single subterm, such as that marked $e_2$ with a new program, such as the $\lambda$-term. The result of such a modification is shown in the following diagram.

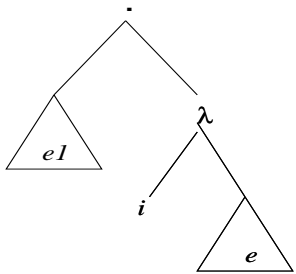

In order to make the idea of intra-program modifications more precise, the term *parameterised program* will be introduced with respect to $E$ and used to define a modification and an extension. A parameterised program $p \in P$ is a term, as defined by $E$, which contains one or more *holes* are denoted by _.

$$P ::= I \mid \lambda I.P \mid P_1 P_2 \mid \_$$

$p(e)$ will mean replacement of all holes in $p$ with the program $e$. A modification to the program $e$ which replaces a subterm $e_1$ with a program $e_2$ is defined with respect to a parameterised program $p$ for which $p(e_1) = e$ and the result is defined as $p(e_2)$. An extension to the program $e$ with respect to the parameterised program $p$ is defined as $p(e)$.

By repeatedly applying modifications and extensions to a program, any new program can be produced. The intra-language changes are not necessarily intended to preserve the meaning of a program, *i.e.* the evaluation of the program before and after the change will not necessarily produce the same results. Such a change will not affect the underlying semantics of the programming language, *i.e.* the machine which defines the meaning of the programs will not change as a result of a modification or an extension. The intra-language changes are intended to reflect the activities which occur at the same level of abstraction, before working out how certain components might be "implemented" in more concrete terms.

Before defining what it means to perform a *port*, it will be necessary to define what it means for a programming language to be consistent and complete with respect to another.

1. *Consistency*[1] between programming languages captures the notion that the two languages will produce similar outcomes from similar programs. The notion of similarity must be made precise and this is usually done by defining it to be a translation from the programs and outcomes of one language to those of the other. Consistency will be defined so that one programming language is consistent with respect to another. If language $P_1$ is consistent with respect to $P_2$ then for some collection of $P_2$ programs there will be a similar collection of $P_1$ programs which produce similar outcomes.

2. *Completeness* with respect to programming languages will be used to mean that consistency follows for all $P_2$ programs.

Different programming languages in the model will all have the same format, *i.e.* $E$, but will differ with respect to the machines which give the languages their different semantics. Each machine will define how to evaluate all programs (although the meaning of some programs with respect to some machines may be undefined) by loading the program onto the machine, performing all possible transitions and then unloading the result. Given a machine $M$ this process will be represented as a function $eval(M) : E \to V$ for some collection of program

---
[1] In this document, when the terms *consistency* and *completeness* are used without qualification, they will be defined to be ... of machines with respect to a translation.

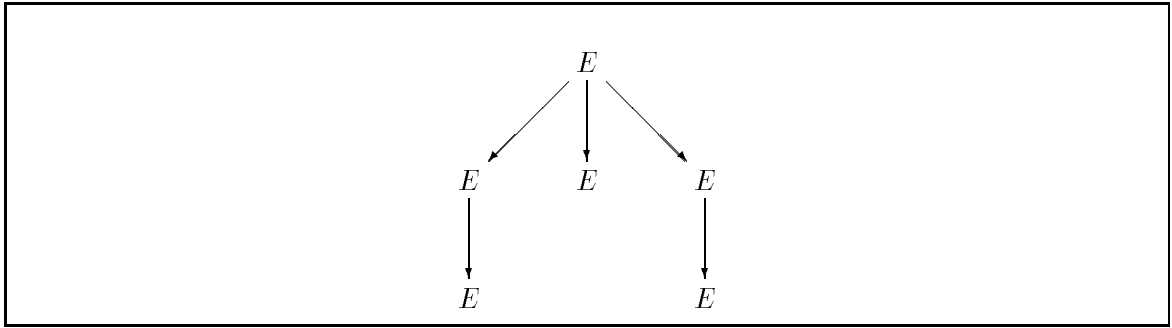

Figure 1: A software development graph

outcomes $V$. Two machines $M_1$ and $M_2$ are equivalent, and therefore consistent with each other, when
$$eval(M_1) \simeq eval(M_2)$$
where $\simeq$ means equal when one or other is defined otherwise undefined. Equivalence is a very strong property between two machines which is not particularly interesting when dealing with ports. More interesting is when some part of the program or the outcome has to be translated in order to guarantee some equivalence. Suppose that there are two possible collections of outcomes $V_1$ and $V_2$ for machines $M_1$ and $M_2$ respectively. Then we can define that the outcomes of machine $M_2$ are consistent with the outcomes of machine $M_1$ if there exists a way of transforming the outcomes of the first into the outcomes of the second, *i.e.* if
$$translateoutcome \circ eval(M_1) \simeq eval(M_2)$$
If the function *translateoutcome* is a total function then we can say that $M_2$ is consistent and *complete* with respect to $M_1$. The same argument holds for translating the programs *before* they are evaluated in order to guarantee some sort of equivalence. Suppose that *translateprogram* is a function which translates from $E$ programs to $E$ programs (this might sound a bit silly, but consider translating a C program with builtin operators for real arithmetic to a C program which inserts all the machinery to do real arithmetic in terms of integer arithmetic), then the evaluation on the machine $M_2$ is consistent with respect to that on machine $M_1$ when
$$eval(M_1) \simeq eval(M_2) \circ translateprogram$$
If the function *translateprogram* is total then the evaluation is both consistent and complete. In general, to show consistency and completeness between two machines it will be necessary to use both types of translation
$$translateoutcome \circ eval(M_1) \simeq eval(M_2) \circ translateprogram$$
A *port* will be a pair of translations between machines for which the target machine is consistent and complete with respect to the source machine such that the translations are *homomorphisms*. This issue is a technical point which guarantees that the translations are defined to be modular and that subterms in source programs are translated to consistent subterms in target programs.

The definitions of *program, machine, calculation, modification, extension, consistency, completeness* and *port* conclude the model of software development. The entire process of software development is represented as a graph, an example of which is shown in figure 1. The nodes of the graph are programs (a.k.a designs and specifications) and the arrows represent changes in terms of modifications, extensions and ports. Where more than one edge leads from a node, this corresponds to alternative possible development paths. Figure 2 shows a single path from the root of the development graph to a current state. The path is a cascade of interchangeable software changes starting with a high level of abstraction and ending up with a more concrete program.

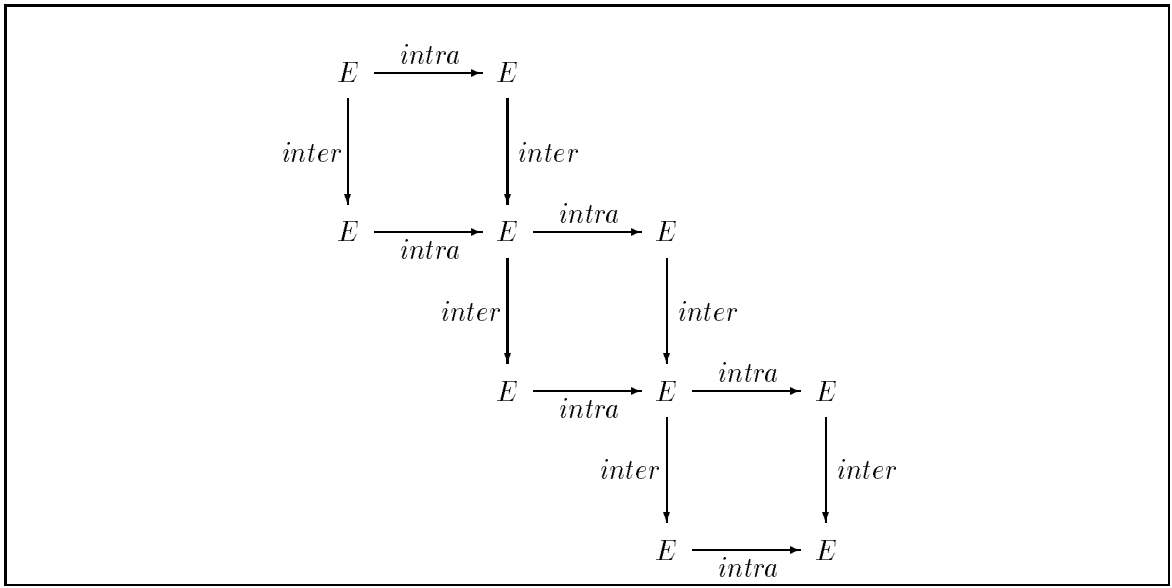

Figure 2: A cascade of software development

## 2 KBS software

AI and KBS software are very difficult to define in any precise manner. The following quotes are taken from a collection of sources which deal with AI and KBS software:

> "[knowledge] goes beyond the notion of information, since it employs a complex structure including, integrating, multiplying and valorizing a great number of information units existing in the brain of an expert or specialist. [ ... ] or whether it is to use this knowledge to accomplish a number of tasks requiring an intellectual endeavour (KBS and expert systems), it is clear that it is necessary to adopt, and may be to rethink, the methodological framework which is dominant in classical software." [1]

> "By AI software we mean software that uses techniques in the field of Artificial Intelligence." [2]

> "We face an initial difficulty in that the notion of AI software is fuzzy – indeed practitioners of AI do not even agree among themselves on what constitutes AI." [2]

> "The class of AI software that we have identified for consideration is often characterised as "knowledge based" meaning that it contains an explicit representation of knowledge about some aspects of the external world." [2]

> "[...]a KBS can generally be described as consisting of 3 major components:
> - a knowledge base.
> - an inference engine.
> - a user interface.
>
> The knowledge base, which contains all the relevant domain knowledge, is a complex entity that must be uniquely developed for each application. However, the inference engine, which navigates the system [...] is a fairly standard mechanism." [3]

"AI is

- a natural extension of computer science and technology.
- a way of producing systems which are more autonomous and resilient than conventional software techniques will allow.
- automated knowledge processing." [4]

"The principal AI languages, Prolog and Lisp, are declarative languages. Note that it is the language design and implementation that determine the flow of control in the program. In this sense AI languages are not essentially sequential in the way that conventional procedural languages are." [4]

"search forms the core of AI based software. Intelligent programs must use efficient search methods." [5]

"We use the term "knowledge based" as a temporary expedient, just as it was once common to talk of digital computers". The term "digital" was later dropped from common speech as the distinction between analog and digital computers became less important. (The same applies to "motor" cars.) "Knowledge based" is simply a transitional term, emphasizing obvious differences between this view [of design] and what has gone on before." [6]

The quotes which are given are typical of the AI and KBS literature when giving overall definitions of these terms. We are interested in any features which characterise KBS software but must be wary of any implementation techniques which are "passing fads" and will be here today and gone tomorrow. We wish to uncover the underlying principles which may be used to construct any of the techniques which are currently used for KBS. We believe that the following is a list of relevant points:

- Stating that AI systems are those which use AI techniques is not very helpful.

- Domain knowledge is generally accepted as essential to KBS but is never really defined. Are text editors knowledge based because they know about ASCII character codes?

- A KBS is often described as consisting of a knowledge base, an inference engine and a user interface. The fact that a system has a user interface is not particularly surprising and certainly not true in all cases (embedded KBS for example). The inference engine is often a form of interpreter for the data values in the knowledge base. Unfortunately, not all KBS have such a construction and therefore knowledge bases and inference engines are not viewed as a characterising feature.

- KBS software is generally called "declarative" which is taken to mean that the programs are less involved with specifying what the individual steps in the computation *do*, as in specifying what large collections of steps will *achieve*. This is a laudable aim and it is certainly not restricted to the area of KBS software. So-called declarative languages represent the leading edge of programming language technology and the fact the KBS software is implemented using such languages is a result of KBS practitioners choosing the most up-to-date tools.

- So-called conventional software is said to use *information* whereas KBS software uses *knowledge* and the difference is that knowledge tends to be more complex and require more involved processing. This is a subjective viewpoint, parsers and compilers were once viewed as being the height of sophistication.

- *Search* is often given as a characterising feature of KBS software. This is a concrete distinction between conventional and KBS software. A search space is a collection of states and transitions between them. A program will search a space until one of a collection of acceptable states is found. Such a program is viewed as being non-deterministic when there is no ordering placed on the development of the search space. Conventional software does not involve search spaces; they are often viewed as a sequenced collection of actions which processes some input to produce some output.

- The search space for a KBS program may be far too large to be fully developed. Even when clever techniques are employed to predict which of a collection of alternative states should be developed next, it is often impossible to guarantee that an acceptable state will be reached given constraints on the program's resources. This will mean that the outcome of a KBS program may be less than optimal. A conventional program will either succeed or fail on a given input; if it succeeds then the answer will be wholly acceptable otherwise the program will completely fail.

From the points given above, we take the two characterising features of KBS software as *non-determinism* which arises due to the search space and *incompleteness* which produces less than optimal outcomes for given inputs due to time and space constraints. These features are fundamental to characterising a KBS system: there may be other features which KBS software typically exhibits, but it is possible for non-KBS software to also have these features.

A *KBS program* will be defined to be an expression in the $\lambda$-calculus which has been extended with a non-deterministic operator, _**or**_, and a construct **fail** which kills off the current computation. The concrete syntax is defined as follows:

$$K ::= I \mid \lambda I.K \mid K_1 K_2 \mid K_1 \text{ or } K_2 \mid \textbf{fail}$$

When an expression $K_1$ **or** $K_2$ is performed, the result will be produced by either $K_1$ or $K_2$. We can view a KBS program as producing all the possible answers and then selecting one of these non-deterministically; in this case, when an **or** expression is performed, both $K_1$ and $K_2$ are performed independently and **both** produce a result for the **or** expression. If **or** expressions are viewed as branching points in the computation then an entire calculation for a KBS program will look like a tree. When a **fail** construct is performed, the current computation is killed off which will mean that the edge on which the **fail** construct lies in the calculation tree will be pruned back to the last branching point.

KBS program semantics will be given semantics in terms of different machines, just as in §1. However, the machines will all have a new distinctive feature which is that the transition relation, $\longmapsto$, will hold between a single machine state and a set of machines states. This captures the non-determinism of KBS programs. As an example, the following is a variation of the SECD

```
let inferenceengine (rules, terminated) data =
    if terminated(data)
    then data
    else lhr (or) (inferenceengine(rules, terminated)) (λ_.fail) (map f apprules)
        where
            apprules = applicablerules(rules, data)
            f(r) = r(data)
```

Figure 3: An inference engine

machine which was defined in §1.

$$(s, b, i :: c, d) \longmapsto \{((b \bullet i) :: s, b, c, d)\}$$

$$(s, b, (\lambda i.k) :: c, d) \longmapsto \{(< i, b, k > :: s, b, c, d)\}$$

$$(s, b, (k_1 k_2) :: c, d) \longmapsto \{(s, b, k_2 :: k_1 :: @ :: c, d)\}$$

$$(< i, b_1, k > :: v :: s, b_2, @ :: c, d) \longmapsto \{([], b_1 \oplus (i \mapsto v), [k], (s, b_2, c, d))\}$$

$$(v :: \_, \_, [], (s, b, c, d)) \longmapsto \{(v :: s, b, c, d)\}$$

$$(s, b, (k_1 \text{ or } k_2) :: c, d) \longmapsto \{(s, b, k_1 :: c, d), (s, b, k_2 :: c, d)\}$$

$$(\_, \_, \textbf{fail} :: \_, \_) \longmapsto \{\}$$

There are also other features which characterise a subclass of KBS software, for example knowledge bases and inference engines. Figure 3 shows a skeleton inference engine (based on [7] pp. 20 − 21) which is constructed from the underlying non-determinism primitive. The inference engine is a program whose inputs are the rules, a predicate which determines when the program has completed and some data. If the program has completed then the data is produced as the outcome. Otherwise there will be a collection of applicable rules. Since only one rule may be applied to the data at any given time, the inference engine will apply *all* the rules non-deterministically.

KBS software development will be defined using the same terms which were described in §1. The essential difference between conventional software development and KBS development is that the ports are not guaranteed to be complete, *i.e.* when a program is translated from one machine to another in order to make the representation more concrete, some of the possible outcomes of the program may be lost. This is not a problem if all of the possible outcomes from a program are acceptable, however in general some of the outcomes from a KBS program will be more acceptable than others. This will mean that after a port, the resulting program may produce less than optimal results. This feature will have a major impact on the quality of KBS software and it is important that the developer has a good idea of the characteristics of the incompleteness so that it is well understood whether or not a change to a program will affect the acceptibility of its outcomes.

Given a machine $M$ and a set of possible outcomes $V$ for a KBS language, the evaluator for $M$ is $eval(M) : K \rightarrow setof(V)$ where the expression $setof(V)$ represents the set of all subsets of $V$. So the evaluator for a KBS programming language differs from that for a conventional language in that the outcome is a set of values rather than just a single value. As before, given two machines $M_1$ and $M_2$ they are equivalent if they produce the same (sets of) outcomes when they evaluate the same programs,

$$eval(M_1) \simeq eval(M_2)$$

Given two sets of outcomes from the machines, $V_1$ and $V_2$ then $M_2$ is consistent with respect to $M_1$ when there is a mapping *translateoutcome* : $V_1 \rightarrow V_2$ which will translate from $M_1$ outcomes to $M_2$ outcomes such that the set of outcomes produced by the machine $M_2$ is a subset of the translated set of outcomes from $M_1$ for the same program. This is expressed by the following inequality:

$$eval(M_2) \subseteq map(translateoutcome) \circ eval(M_1)$$

Notice that this differs from the notion of conventional software consistency in that $M_2$ need only produce a subset after a translation. Alternatively, a program may be translated before being evaluated. Suppose that *translateprogram* : $K \rightarrow K$ is a program translation, then $M_2$ is consistent with $M_1$ when the set of outcomes produced by $M_1$ is a superset of that produced by $M_2$ after a translation,

$$eval(M_2) \circ translateprogram \subseteq eval(M_1)$$

Finally, KBS language consistency is defined in terms of both an outcome and a program translation,

$$eval(M_2) \circ translateprogram \subseteq map(translateoutcome) \circ eval(M_1)$$

In general, a port will be used in software development to translate from an idealised programming language to a more concrete language. The idealised language may be unfettered by resource constraints such as memory usage and execution duration. When conventional languages are ported, the program is intended to do exactly the same thing before and after the port; the difference is that the calculation which is performed after the port will be in terms of elements which have more "implementation detail". When KBS languages are ported, the effect of filling in "implementation detail" goes hand in hand with the problems of finite resources which is why the collection of outcomes produced by the program after the port will be a subset of those before the port. This is a characteristic feature of KBS software development and is a genuine difference between KBS and conventional software. The impact of this feature on software quality is that the loss of completeness due to design and implementation decisions can lead to programs which produce unacceptable outcomes, take too long to produce outcomes and have memory useage problems. By understanding the nature of KBS *evaluation* it is possible to take these issues into account early on in the specification, design and implementation of the system and thereby reduce the risk of producing an unacceptable product.

The view of KBS software which is given in this section can be used as a basis for KBS analysis. The MOSES project intends to produce recommendations for KBS development which will guarantee a level of quality in military systems. A suggested workplan is to identify all of the techniques which are involved in military KBS (*eg.* data fusion, planning, classification, pattern matching *etc.*) and to give an idealised representation for each technique (by modifying and extending the KBS language given here). By analysing each feature, the elements which affect the quality of a product which employs the feature can be identified. These activities will lead to a precise description of the issues which affect the quality in military systems which employ KBS techniques. This will provide the project with the required information in order to produce a list of guidelines with respect to the procurement, specification, design and implementation of military KBS systems.

## 3 Quality terms

This section will take quality terms which have been used in the milestone report [8] and relate them to the models of conventional and KBS software which have been described in this document. Each term is given a brief description which places it in the context of this document.

**Integration** Integration is defined to be the process of composing two different software systems together to form a new software system. In the extreme case, both software systems will have different semantics and will be parameterised programs where the other program is supplied as the parameter. Quality will be affected by the compatibilty of the two software systems *i.e.* by the existence and completeness of a translation from one of the language systems to the other. In addition to performing computations with respect to an interface, the resulting system may involve *merging* the two original systems in which case quality will be affected by the consistency of the merge (*i.e.* whether or not the merge leaves the resulting system in a sensible state).

**Porting** Porting has already been defined for both conventional and KBS software. KBS quality is affected by the completeness of the port. In general, the quality of a port will be affected by the modularity of the translation from one language to another. An idealised port will be a homomorphism from the syntactic constructs of one language to those of another. The more languages differ, the less likely the translation will be a homomorphism and the greater the difficulty of ensuring that the port is a success.

**Completeness** Completeness must be defined with respect to something which is defined to be complete. For example a "knowledge base" $K$ may be defined to be complete with respect to all the known facts $F$ if there exists a mapping which generates all the elements of $F$ using just the elements of $K$. A translation from one programming language to another is complete if it is total. KBS software has incompleteness built in as a characterising feature. It is important that the nature of the incompleteness is understood in order that quality is maintained.

**Consistency** Consistency is a relation between two or more things. The consistency of a software modification is measured by showing that the software can be shown to do roughly the same things after the modification (where *roughly* is captured as a mapping on outcomes and programs) as it did before. A "knowledge base" $K$ may be said to be consistent with respect to the set of all known facts $F$ if all of the deductions which can be made from the data in $K$ are true facts in $F$ (*i.e.* no incorrect deductions can be made). The more inconsistencies which creep into a system (development or execution) the more likely the system will be to fail; quality is maintained by ensuring some degree of consistency. Typically a quality check will be "Given a modification of type $X$ the system will be consistent with respect to $Y$" which shows that nothing has gone wrong due to $X$.

**Testability** Testability refers to the property of a system which enables its behaviour to be compared against expected results. This is a quality issue because the higher the degree of testability (and number of tests performed) the greater the likelihood of the system performing as expected. There are three things which will affect the degree of testability which a system exhibits:

1. A description of all possible input data.
2. A description of the calculations which a system performs on all inputs.
3. A description of the required results from all possible inputs.

Using these descriptions, it will be possible to show (to the desired level of detail) that the correct behaviour is achieved by the system. KBS programs are often described as being difficult to test because their behaviour is unpredictable, *eg*

> "The problems addressed by AI-software are generally somewhat ill-defined, and a clear statement of requirements for the task the software is to perform is often lacking. This means that the notions of success and failure are vague, and evaluation is correspondingly difficult. In addition, the heuristic techniques employed in AI software tend to render it fragile, or unstable: very similar inputs may produce wildly different outputs. This makes extrapolation from behavior on test cases very risky."[2]

There are two issues here. The first, and most important, is that AI-software is "somewhat ill-defined". Whilst this may be true of many AI artifacts, there seems no intrinsic property of AI-related problems which makes them impossible to specify precisely. The ill-defined nature may be related to the incompleteness of the KBS (or AI) system, where the desired solution is the complete one and the problem specification is along the lines of "The software will perform task X and *squeeze as much completeness out of the implementation as possible using the resources available.*" It may be this open ended type of specification which leads to ill-defined specifications because it tends to say nothing about what the system will do *eg* "Advise on the deployment of countermeasures in an air to sea attack scenario and *produce the best advice possible*." This is a serious problem, but it is of primary importance with respect to achieving desired quality standards and deserves further consideration within the MOSES project.

The second issue is related to incompleteness but cannot really be taken seriously, either a collection of rules (*heuristic* usually means rules of some kind) means something or they mean nothing, there are no half measures!

**Modifiability** The term *modify* with respect to software has been defined in this document. It refers to the activity of replacing a subterm of the abstract syntax tree. Modifiability is an ill-defined software property which relates in some way to the act of modification. One desirable property of software is that it is modular; the software can be modified locally to achieve some change in behaviour, without having to affect other areas. A program development is *completely modular* when all intra-language changes lead to legal programs. Modifiability is a measure of the modularity of the program development. (*Hmm – not sure about that*)

**Extensibility** The term *extension* with respect to software has been defined in this document. It refers to the activity of wrapping a parameterised program around an existing program.

**Maintainability** Maintenance refers to the activity by which a system is modified in order to change some undesirable feature of its behaviour. The change may be because the system does not achieve the correct results or because the correct results are achieved in an unacceptable way. A software system is maintainable when it is both modifiable and extensible.

**Robustness** The robustness of a software system is a measure of how well it deals with illegal input data. Input data is illegal when it does not conform to the description given in the specification (see testability above). Because of incompleteness, KBS software must be shown to be robust in the absence of acceptable results.

These descriptions have been given in very broad terms because of the generality of the definitions of conventional and KBS software. If the model of KBS software is made more specific by analysing a particular KBS technique, then the descriptions will be tailored to the technique and the issues involved can be given in more detail. For example, nothing has been said about memory usage *vs.* execution time *vs.* quality; however the KBS software model can be extended with information about time and space and various KBS techniques analysed with respect to the new model and whether or not quality standards can be achieved.